\documentclass[aps,preprint]{revtex4}%
\usepackage{amsfonts}
\usepackage{amsmath}
\usepackage{amssymb}
\usepackage{graphicx}%
\providecommand{\U}[1]{\protect\rule{.1in}{.1in}}

\begin{document}
\preprint{ }
\title[Beyond the Dirac phase factor]{Beyond the Dirac phase factor}
\author{Konstantinos Moulopoulos}
\email{cos@ucy.ac.cy}
\affiliation{University of Cyprus, Department of Physics, 1678 Nicosia, Cyprus}

\begin{abstract}
We report on previously overlooked solutions of the usual gauge transformation
equations that exhibit a new form of nonlocal quantal behavior with the
well-known Relativistic Causality of classical fields affecting directly the
phases of wavefunctions. The new nonlocalities compete with Aharonov-Bohm
behaviors and they provide$\boldsymbol{:}$ a correction to a number of
erroneous results in the literature, a new interpretation of semiclassical
observations and further extensions to delocalized states, a natural remedy of
earlier \textquotedblleft paradoxes\textquotedblright, and a new formulation
in the study of time-dependent slit-experiments.

\end{abstract}
\volumeyear{2010}
\volumenumber{number}
\issuenumber{number}
\eid{identifier}
\date[October 15, 2010]{}
\startpage{1}
\endpage{ }
\maketitle

\section{\bigskip Introduction}

The Dirac phase factor $-$ with a phase containing integrals over potentials
(of the general form $%
{\displaystyle\int\limits^{x}}
\boldsymbol{A}\cdot dx^{\prime}-c%
{\displaystyle\int\limits^{t}}
\phi dt^{\prime}$) $-$ is the standard and widely used solution of the usual
gauge transformation equations of Electrodynamics (with $\boldsymbol{A}$ and
$\phi$ vector and scalar potentials respectively). In a quantum mechanical
context, it connects wavefunctions of two systems (with different potentials)
that experience the same classical fields, i.e. either systems that are
gauge-equivalent (a trivial case with no physical consequences), or systems
that exhibit phenomena of the Aharonov-Bohm type (magnetic or electric) $-$
and then this Dirac phase has nontrivial observable consequences. However, it
has not been realized that the gauge transformation equations, viewed in a
more general context, can have \textit{more general solutions} than simple
Dirac phases, and these lead to wavefunction-\textit{phase-nonlocalities} that
have been widely overlooked and that seem to have important physical
consequences. In this paper we will briefly demonstrate these generalized
solutions and will present cases (and closed analytical results for the
wavefunction-phases) that actually connect (or map) two quantal systems that
are \textbf{neither physically equivalent nor of the usual Aharonov-Bohm
type}. We will also explore the consequences of the new (\textit{nonlocal})
contributions (that appear in the wavefunction-phases) and will see that they
are numerous and important; they are also of a different type in static and in
time-dependent configurations (and in the latter cases they seem to lead to
Relativistically \textit{causal} behaviors, that apparently resolve earlier
\textquotedblleft paradoxes\textquotedblright\ arising in the literature from
the use of standard Dirac phase factors).

Let us first remind the reader of a property that is more general than usually
realized$\boldsymbol{:}$ the solutions $\Psi(\mathbf{r},t)$ of the
$t$-dependent Schr\"{o}dinger (or Dirac) equation (SE) for a quantum particle
of charge $q$ that moves (as a test particle) in two distinct sets of
(predetermined and classical) vector and scalar potentials ($\mathbf{A}%
_{1},\phi_{1}$) and ($\mathbf{A}_{2},\phi_{2}$), that are generally spatially-
and temporally-dependent [and such that, at the spacetime point of observation
$(\mathbf{r},t)$, the magnetic and electric fields are the same in the two
systems], are formally connected through%

\begin{equation}
\Psi_{2}(\mathbf{r},t)=e^{i\frac{q}{\hbar c}\Lambda(\mathbf{r},t)}\Psi
_{1}(\mathbf{r},t), \label{Basic1}%
\end{equation}
with the function $\Lambda(\mathbf{r},t)$ required to satisfy%

\begin{equation}
\nabla\Lambda(\mathbf{r,t})=\mathbf{A}_{2}(\mathbf{r},t)-\mathbf{A}%
_{1}(\mathbf{r},t)\qquad and\qquad-\frac{1}{c}\frac{\partial\Lambda
(\mathbf{r},t)}{\partial t}=\phi_{2}\left(  \mathbf{r},t\right)  -\phi
_{1}(\mathbf{r},t). \label{Basic11}%
\end{equation}
The above property can be immediately proven by substituting each $\Psi_{i}$
into its corresponding ($i_{th}$) time-dependent SE (namely with the set of
potentials ($\mathbf{A}_{i}(\mathbf{r},t),\phi_{i}(\mathbf{r},t)$%
))$\boldsymbol{:}$ one can then easily see that (\ref{Basic1}) and
(\ref{Basic11}) guarantee that both SEs are indeed satisfied together (after
cancellation of a global phase factor in system 2). [In addition, the equality
of all classical fields at the observation point, namely $\mathbf{B}%
_{2}(\mathbf{r},t)=\nabla\times\mathbf{A}_{2}(\mathbf{r},t)=\nabla
\times\mathbf{A}_{1}(\mathbf{r},t)=$ $\mathbf{B}_{1}(\mathbf{r},t)$ for the
magnetic fields (MFs) and $\mathbf{E}_{2}(\mathbf{r},t)=-\nabla\phi_{2}\left(
\mathbf{r},t\right)  -\frac{1}{c}\frac{\partial\mathbf{A}_{2}(\mathbf{r}%
,t)}{\partial t}=-\nabla\phi_{1}\left(  \mathbf{r},t\right)  -\frac{1}{c}%
\frac{\partial\mathbf{A}_{1}(\mathbf{r},t)}{\partial t}=\mathbf{E}%
_{1}(\mathbf{r},t)$ for the electric fields (EFs), is obviously consistent
with all equations (\ref{Basic11}) $-$ provided, at least, that $\Lambda
(\mathbf{r},t)$ is such that interchanges of partial derivatives with respect
to all spatial and temporal variables (at the point $\left(  \mathbf{r}%
,t\right)  $) are allowed].

The above fact is of course well-known within the framework of the theory of
quantum mechanical gauge transformations (the usual case being with
$\mathbf{A}_{1}=\phi_{1}=0$, hence a mapping from a system with no
potentials)$\boldsymbol{;}$ but in that framework, these transformations are
supposed to connect (or map) two \textit{physically equivalent systems} (more
rigorously, this being true for ordinary gauge transformations, in which case
the function $\Lambda(\mathbf{r},t)$, the so-called gauge function, is unique
(single-valued) in spacetime coordinates). In a formally similar manner, the
above argument is also often used in the context of the so-called
\textquotedblleft singular gauge transformations\textquotedblright, where
$\Lambda$ is multiple-valued, but the above equality of classical fields is
still imposed (at the observation point, that always lies in a physically
accessible region)$\boldsymbol{;}$ then the above simple phase mapping (at all
points of the physically accessible spacetime region, that experience equal
fields) leads to the standard phenomena of the Aharonov-Bohm (AB) type,
reviewed below, where \textit{unequal fields in physically-inaccessible
regions} have observable consequences. However, we should keep in mind that
that above property ((\ref{Basic1}) and (\ref{Basic11}) taken together) can be
\textit{more generally valid} $-$ and, as already stated, we will present
cases (and closed analytical results for the appropriate phase function
$\Lambda(\mathbf{r},t)$) that actually connect (or map) two systems (in the
sense of (\ref{Basic1})) that are \textbf{neither physically equivalent nor
exhibiting the usual AB behaviors}. And naturally, because of the above
provision of field equalities at the observation point, it will turn out that
any nonequivalence of the two systems will involve \textit{remote} regions of
spacetime, namely regions that do \textit{not} contain the observation point
$(\mathbf{r},t)$ (and in which regions, as we shall see, the classical fields
experienced by the particle may be \textit{different} in the two systems).

Returning to the standard cases, usual $\Lambda$'s are given in terms of Dirac
phases, namely integrals over potentials. I.e. in static cases, and if, for
simplicity, we start from system 1 being completely free of potentials
($\mathbf{A}_{1}=\phi_{1}=0$), the wavefunctions of the particle in system 2
(moving only in a static vector potential $A(\mathbf{r})$) will acquire an
extra phase with an appropriate \textquotedblleft gauge
function\textquotedblright\ $\Lambda(\mathbf{r})$ that must satisfy$\ \nabla
\Lambda(\mathbf{r})=\mathbf{A}(\mathbf{r}).$ The standard (and widely-used)
solution of this is the line integral $\Lambda(\mathbf{r})=\Lambda
(\mathbf{r}_{\mathbf{0}})+\int_{\mathbf{r}_{0}}^{\mathbf{r}}\mathbf{A}%
(\mathbf{r}^{\prime})\boldsymbol{.}d\mathbf{r}^{\prime}$ (which, by
considering two paths encircling an enclosed inaccessible magnetic flux,
formally leads to the well-known magnetic AB effect\cite{AB}). It should
however be stressed that the above is only true if $\nabla\Lambda
(\mathbf{r})=\mathbf{A}(\mathbf{r})$ is valid for \textbf{all} points
$\mathbf{r}$ of the region where the particle moves, i.e. if the particle in
system 2 moves (as a narrow wavepacket) always outside MFs ($\nabla
\times\mathbf{A}=0$ everywhere). Similarly, if the particle in system 2 moves
only in a spatially uniform scalar potential $\phi(t)$, the appropriate
$\Lambda$ must satisfy $-\frac{1}{c}\frac{\partial\Lambda(t)}{\partial t}%
=\phi(t),$ the standard solution being $\Lambda(t)=\Lambda(t_{0})-c\int%
_{t_{0}}^{t}\phi(t^{\prime})dt^{\prime}$ that gives the extra phase acquired
by system 2 (this result formally leading to the electric AB effect\cite{AB}
by applying it to two equipotential regions, such as two metallic cages held
in distinct time-dependent scalar potentials). Once again however it should be
stressed that the above is only true if $-\frac{1}{c}\frac{\partial\Lambda
(t)}{\partial t}=\phi(t)$\ is valid at \textbf{all} times $t$ of interest,
i.e. if the particle in system 2 moves (as a narrow wavepacket) always outside
EFs ($\mathbf{E}=-\nabla\phi-\frac{1}{c}\frac{\partial\mathbf{A}}{\partial
t}=0$ \ at all times). (In the electric AB setup, the above is ensured by the
fact that $t$ lies in an interval of a finite duration $T$\ for which the
potentials are turned on, in combination with the narrowness of the
wavepacket$\boldsymbol{;}$ and the assumption is that, during $T$, the
particle has vanishing probability of being at the edges of the cage where the
potential starts having a spatial dependence. The reader is referred to
Appendix B of Peshkin\cite{Peshkin} that demonstrates the intricasies of the
electric AB effect, to which we return with an important comment at the end of
this paper).

For potentials more general than in the above cases, (and if, for notational
simplicity, we restrict our attention to only one spatial variable
$x$)$\boldsymbol{\ }$it is usually stated that the general gauge function that
connects (through a phase factor $e^{i\frac{q}{\hbar c}\Lambda(x,t)}$) the
wavefunctions of a quantum system with no potentials to those in a general set
$(\boldsymbol{A},\phi)$ is the obvious combination (and a natural extension)
of the above two forms, namely%

\begin{equation}
\Lambda(x,t)=\Lambda(x_{0},t_{0})+%
{\displaystyle\int\limits_{x_{0}}^{x}}
\boldsymbol{A(}x^{\prime},t)\cdot d\boldsymbol{x}^{\prime}-c%
{\displaystyle\int\limits_{t_{0}}^{t}}
\phi(x,t^{\prime})dt^{\prime}, \label{wrong}%
\end{equation}
which, however, is generally \textbf{incorrect } for $x$ and $t$ uncorrelated
variables$\boldsymbol{:}$ it does \textbf{not }generally\textbf{ }satisfy the
standard system\textbf{\ }(\ref{Basic11}) (viewed as a system of partial
differential equations (PDEs)), namely%

\begin{equation}
\nabla\Lambda=\mathbf{A\qquad and\qquad-}\frac{1}{c}\frac{\partial\Lambda
}{\partial t}=\phi. \label{BasicPDE}%
\end{equation}
Indeed$\boldsymbol{:}$ (i) When the $\nabla$ operator acts on eq.(\ref{wrong}%
), it gives the correct $A(x,t)$ from the 1st term, but it also gives some
annoying additional nonzero quantity from the 2nd term (that survives because
of the $x$-dependence of $\phi$); hence it invalidates the first of the basic
system (\ref{BasicPDE}). (ii) Similarly, when the $\mathbf{-}\frac{1}{c}%
\frac{\partial}{\partial t}$ operator acts on eq.(\ref{wrong}), it gives the
correct $\phi(x,t)$ from the 2nd term, but it also gives some annoying
additional nonzero quantity from the 1st term (that survives because of the
$t$-dependence of $\mathbf{A}$); hence it invalidates the second of the basic
system (\ref{BasicPDE}). It is only when $\mathbf{A}$ is $t$-independent, and
$\phi$ is spatially-independent, that eq.(\ref{wrong}) is correct. It is also
interesting to note that the line integrals appearing in (\ref{wrong}) do
\textit{not} form a path (in spacetime) that connects the initial to the final
point (see below). [An alternative form that is also given in the literature
is again eq.(\ref{wrong}), but with the variables that are not integrated over
implicitly assumed to belong to the initial point (hence a $t_{0}$ replaces
$t$ in $\boldsymbol{A}$, and an $x_{0}$ replaces $x$ in $\phi$). However, one
can see again that the system (\ref{BasicPDE}) is \textit{not} satisfied (the
above differential operators, when acted on $\Lambda$, give $\boldsymbol{A}%
(x,t_{0})$ and $\phi(x_{0},t)$, hence \textit{not} the values of the
potentials at the point of observation $(x,t)$ as they should), \textit{this
not being an acceptable solution either}. And in this case also there is no
spacetime-path connecting the initial $(x_{0},t_{0})$ to the final point
$(x,t)$ either, as the reader can easily verify]. What is the problem here,
or, better put, what is the deeper reason for the above inconsistency? The
short answer is the uncritical use of Dirac phase factors that come from
path-integral treatments (where $x$ and $t$ are not \textbf{uncorrelated}
variables, but actually correlated to produce a path $x(t)$). The general
inadequacy of (\ref{wrong}) was actually one of the main points that has
motivated this work. By looking for the most general form of $\Lambda$ that
solves the basic system of PDEs we have recently found \textit{generalized
results} that actually \textit{correct} eq.(\ref{wrong}) in 2
ways$\boldsymbol{:}$ through the proper appearance of $x_{0}$ and $t_{0}$ (as
in eq.(\ref{LambdaStatic1}) and eq.(\ref{LambdaStatic2}) of next Section) $-$
which happens to give a path-sense (that connects the initial to the final
point) in either of the two solutions (see Fig.1), being therefore consistent
with Feynman's path integral result in the special case of narrow-wavepacket
states $-$ but most importantly, through the additional presence of
\textit{novel nonlocal terms} that had so far been overlooked\cite{jphysa}.
These generalized results are the \textbf{exact} solutions of the system
(\ref{BasicPDE}) but, even most importantly, the formulation (and methodology
of solution) that produces them, if applied to $\Lambda(x,y)$ (in the 2-D
static case) and also to $\Lambda(x,y,t)$ (in the full dynamical 2-D case),
leads to the exact (nontrivial) forms of the phase function $\Lambda$ that,
apart from satisfying (in all cases) the system (\ref{BasicPDE}), seems to
also have far reaching consequences for the wavefunction-phases in the
Schr\"{o}dinger picture (the most important being their causal behavior).

Summarizing, we will see in this paper that the full form of a general
$\Lambda$ goes beyond the usual Dirac phases$\boldsymbol{:}$ apart from
integrals over potentials, it also generally contains terms of classical
fields that act \textit{nonlocally} (in spacetime) on the solutions of the
$t$-dependent SE. As a result, the phases of wavefunctions in the
Schr\"{o}dinger picture are affected nonlocally by MFs\ and EFs $-$ nonlocal
contributions that have apparently escaped from path-integral approaches. We
will then focus on two types of application of the new
formulation$\boldsymbol{:}$ (i) Application to particles passing through
static MFs or EFs will lead to cancellations of AB phases at the observation
point; these cancellations will be linked to behaviors at the semiclassical
level (to early experimental observations by Werner \& Brill or to recent
reports of Batelaan \& Tonomura) but will be shown to be far more general
(valid not only for narrow wavepackets but also for completely delocalized
quantum states). By using them we will provide a new interpretation of
semiclassical results and we will point out a number of sign errors in popular
reports in the literature$\boldsymbol{:}$ we will clearly show that
semiclassical phase-differences picked up by classical trajectories (deflected
by fields) are \textit{opposite} (and \textit{not} equal, as usually stated or
implied) to the corresponding \textquotedblleft AB phase\textquotedblright%
\ (due to the flux enclosed by the same trajectories). (ii) Application to
$t$-dependent situations will provide a remedy for a number of misconceptions
(on improper use of simple Dirac phase factors) propagating in the literature
(Feynman, Erlichson and others), and will lead to nontrivially extended phases
that contain an AB part and a nonlocal field-part$\boldsymbol{:}$ their
competition will be shown to recover Relativistic Causality in earlier
\textquotedblleft paradoxes\textquotedblright\ (such as the van Kampen
thought-experiment) and will provide a fully quantitative formulation of
Peshkin's qualitative discussion (on expected causal behavior) in the electric
AB effect (discussion that was also based on a simple Dirac phase factor). The
temporal nonlocalities found in this work demonstrate in part \textit{a causal
propagation of phases of quantum wavefunctions in the Schr\"{o}dinger picture}
(through the well-known causal propagation of fields), something that may open
a new and direct way for addressing $t$-dependent double-slit experiments and
the associated causal issues.

\section{1-D Dynamic case}

Let us first consider 1-D cases and find the proper $\Lambda(x,t)$ that takes
us from (maps) a system in a set ($A_{1}$,$\phi_{1}$) to a set ($A_{2}$%
,$\phi_{2}$). As already emphasized, we must assume that at the point $(x,t)$
of observation we have equal EFs, i.e. $-\frac{\partial\phi_{2}}{\partial
x}-\frac{1}{c}\frac{\partial A_{2}}{\partial t}=-\frac{\partial\phi_{1}%
}{\partial x}-\frac{1}{c}\frac{\partial A_{1}}{\partial t}$, but we will
\textit{not} exclude the possibility of the two systems passing through
\textit{different }EFs in other regions of spacetime (that do \textit{not}
contain the observation point). In fact, this possibility will \textit{come
out naturally} from a careful solution of the basic PDEs, namely$\ \frac
{\partial\Lambda}{\partial x}=A,$ $-\frac{1}{c}\frac{\partial\Lambda}{\partial
t}=\phi.$ This system is underdetermined in the sense that we only have
knowledge of $\Lambda$ at an initial point $(x_{0},t_{0})$ and with no further
boundary conditions (hence multiplicities of solutions being generally
expected, see below). By following a careful procedure of
integrations\cite{jphysa} we finally obtain 2 distinct solutions (depending on
which eq. we integrate first)$\boldsymbol{:}$ the first solution is%

\begin{equation}
\Lambda(x,t)=\Lambda(x_{0},t_{0})+%
{\displaystyle\int\limits_{x_{0}}^{x}}
A(x^{\prime},t)dx^{\prime}-c%
{\displaystyle\int\limits_{t_{0}}^{t}}
\phi(x_{0},t^{\prime})dt^{\prime}+\left\{  c%
{\displaystyle\int\limits_{t_{0}}^{t}}
dt^{\prime}%
{\displaystyle\int\limits_{x_{0}}^{x}}
dx^{\prime}E(x^{\prime},t^{\prime})+g(x)\right\}  +\tau(t_{0})
\label{LambdaStatic1}%
\end{equation}
with$\ g(x)$ required to be chosen so that the quantity$\ \left\{  c%
{\displaystyle\int\limits^{t}}
{\displaystyle\int\limits^{x}}
E+g(x)\right\}  $\ is indep. of $x$, and the second solution is%

\begin{equation}
\Lambda(x,t)=\Lambda(x_{0},t_{0})+%
{\displaystyle\int\limits_{x_{0}}^{x}}
A(x^{\prime},t_{0})dx^{\prime}-c\int_{t_{0}}^{t}\phi\left(  x,t^{\prime
}\right)  dt^{\prime}+\left\{  -c%
{\displaystyle\int\limits_{x_{0}}^{x}}
dx^{\prime}%
{\displaystyle\int\limits_{t_{0}}^{t}}
dt^{\prime}E(x^{\prime},t^{\prime})+\hat{g}(t)\right\}  +\chi(x_{0})
\label{LambdaStatic2}%
\end{equation}
with$\ \hat{g}(t)$ to be chosen in such a way that$\ \left\{  -c%
{\displaystyle\int\limits^{x}}
{\displaystyle\int\limits^{t}}
E+\hat{g}(t)\right\}  \ $is indep. of $t$. We can directly verify that
(\ref{LambdaStatic1}) or (\ref{LambdaStatic2}) are indeed solutions of the
basic PDEs. [For (\ref{LambdaStatic1}) we have (even for $E(x^{\prime
},t^{\prime})\neq0$)$\boldsymbol{:}$ $\frac{\partial\Lambda(x,t)}{\partial
x}=A(x,t)$ satisfied trivially (because $\left\{  ..\right\}  $ is indep. of
$x$), and \ $-\frac{1}{c}\frac{\partial\Lambda(x,t)}{\partial t}=-\frac{1}{c}%
{\displaystyle\int\limits_{x_{0}}^{x}}
\frac{\partial A(x^{\prime},t)}{\partial t}dx^{\prime}+\phi(x_{0},t)-%
{\displaystyle\int\limits_{x_{0}}^{x}}
E(x^{\prime},t)dx^{\prime}$, and then with the substitution $-\frac{1}{c}%
\frac{\partial A(x^{\prime},t)}{\partial t}=\frac{\partial\phi(x^{\prime}%
,t)}{\partial x^{\prime}}+E(x^{\prime},t)$ we obtain $\ -\frac{1}{c}%
\frac{\partial\Lambda(x,t)}{\partial t}=%
{\displaystyle\int\limits_{x_{0}}^{x}}
\frac{\partial\phi(x^{\prime},t)}{\partial x^{\prime}}dx^{\prime}+%
{\displaystyle\int\limits_{x_{0}}^{x}}
E(x^{\prime},t)dx^{\prime}+\phi(x_{0},t)-%
{\displaystyle\int\limits_{x_{0}}^{x}}
E(x^{\prime},t)dx^{\prime}$. Since the 2nd and 4th terms \textit{cancel each
other}, and the 1st term is \ $%
{\displaystyle\int\limits_{x_{0}}^{x}}
\frac{\partial\phi(x^{\prime},t)}{\partial x^{\prime}}dx^{\prime}%
=\phi(x,t)-\phi(x_{0},t)$ we obtain $-\frac{1}{c}\frac{\partial\Lambda
(x,t)}{\partial t}=\phi(x,t).$ $\checkmark$ \ We have directly shown therefore
that the basic system of PDEs is indeed satisfied by our \textit{generalized}
solution (\ref{LambdaStatic1}) \textbf{even for any nonzero} $E(x^{\prime
},t^{\prime})$ \ (in regions $(x^{\prime},t^{\prime})\neq(x,t)$). (Note
however that at the point of observation $E(x,t)=0$, signifying the essential
fact that the fields in the two systems are identical (recall that
$E=E_{2}-E_{1}$) at the point of observation $(x,t)$). It can similarly be
shown that (\ref{LambdaStatic2}) is also a solution]. In (\ref{LambdaStatic1})
and (\ref{LambdaStatic2}) the placement of $x_{0}$ and $t_{0}$ gives a
\textquotedblleft path-sense\textquotedblright\ to the line integrals in each
solution (each path consisting of 2 perpendicular line segments connecting
$(x_{0},t_{0})$ to $(x,t),$ with solution (\ref{LambdaStatic1}) having a
clockwise and solution (\ref{LambdaStatic2}) a counter-clockwise sense, see
red and green arrow paths in Fig.1)$\boldsymbol{;\ }$this way a natural
\textit{rectangle} is formed, within which the enclosed \textquotedblleft
electric fluxes\textquotedblright\ in spacetime appear to be crucial (showing
up as nonlocal contributions of the EFs-difference from regions $(x^{\prime
},t^{\prime})$ of space and time \textit{that are remote to the observation
point} $(x,t)$). These nonlocal terms in $\Lambda$ have a direct effect on the
wfs' phases at $(x,t)$. The actual manner in which this happens is determined
by the functions$\ g(x)$ or$\ \hat{g}(t)$--these must be chosen in such a way
that they satisfy their respective conditions. In Fig.1a we show an
extended\textit{ vertical} striped-$E$-distribution (the case of a 1-D
capacitor that is arbitrarily charged for all time), where, for $x$ located
outside (and on the right of) the capacitor, the simplest proper choices are
$g(x)=0$ and $\hat{g}(t)=+c%
{\displaystyle\int\limits^{x}}
{\displaystyle\int\limits^{t}}
E$ (since the quantity $%
{\displaystyle\int\limits^{t}}
{\displaystyle\int\limits^{x}}
E$ is already indep. of $x$\ (a displacement of the $(x,t)$-corner of the
rectangle to the right does not change the enclosed \textquotedblleft electric
flux\textquotedblright\ $-$ hence the choice of $g(x)=0$) but \textit{is not a
constant}$\boldsymbol{:}$ this enclosed flux depends on $t$ (since it
\textit{does change} with a displacement of the $(x,t)$-corner upwards) -
hence the choice of $\hat{g}(t)$ above). These choices then of $g(x)$ and
$\hat{g}(t)$ lead (through (\ref{LambdaStatic1}) and (\ref{LambdaStatic2})) to
new (generalized) solutions for this particular field-configuration. We then
note that the \textit{difference} of the two solutions (\ref{LambdaStatic1})
and (\ref{LambdaStatic2}) is \textit{zero }(the flux determined by the
potential-integrals is exactly cancelled by the nonlocal term of EFs), a
cancellation effect that is important and that will be generalized below. For
other shapes of $E$ the choices of $g(x)$ and $\hat{g}(t)$ will be
different$\boldsymbol{:}$ for an extended \textit{horizontal} strip (the case
of a nonzero EF in all space that has a finite duration $T$), proper choices
(for observation instant $t>T$) are $\hat{g}(t)=0$ \ and $g(x)=-c%
{\displaystyle\int\limits_{t_{0}}^{t}}
{\displaystyle\int\limits_{x_{0}}^{x}}
E$ (since the electric flux enclosed in the \textquotedblleft observation
rectangle\textquotedblright\ now depends on $x$, but not on $t$) $-$ or a more
involved example would correspond to a triangular shape (see Fig.1b for the
corresponding magnetic case to be discussed later), where the enclosed flux
depends on \textit{both} $x$ and $t$ (but can be shown to be separable, see
next Section). As for the last constant terms $\tau(t_{0})$ and $\chi(x_{0})$
(what we will call \textquotedblleft multiplicities\textquotedblright), these
are only present when $\Lambda$ is expected to be multivalued, i.e. in cases
of motion in multiple-connected spacetimes, and are then related to the fluxes
in the inaccessible regions: in the electric AB setup, the prototype of
multiple-connectivity in spacetime, it turns out\cite{jphysa} that $\tau
(t_{0})=-\chi(x_{0})=$ enclosed \textquotedblleft electric
flux\textquotedblright, and if these values are substituted in
(\ref{LambdaStatic1}) and (\ref{LambdaStatic2}) they cancel out the new
nonlocal terms and lead to the usual electric AB result. In simple-connected
spacetimes, it can be rigorously shown\cite{jphysa} that solutions
(\ref{LambdaStatic1}) and (\ref{LambdaStatic2}) are equal (with $g(x)$ being
equal to the $t$-indep. bracket of (\ref{LambdaStatic2}), and $\hat{g}(t)$
being equal to the $x$-indep. bracket of (\ref{LambdaStatic1})), the nonlocal
terms having therefore the tendency to exactly cancel the \textquotedblleft AB
terms\textquotedblright\ (this being true \textit{for arbitrary shapes and
analytical form of }$E(x,t)$).

\section{2-D Static Case}

\bigskip The same method applied to static 2-D cases (now for the system of
PDEs $\frac{\partial\Lambda}{\partial x}=A_{x},$ $\frac{\partial\Lambda
}{\partial y}=A_{y}$) finally gives 2 general solutions\cite{jphysa}%
$\boldsymbol{:}$ the first is%

\begin{equation}
\Lambda(x,y)=\Lambda(x_{0},y_{0})+%
{\displaystyle\int\limits_{x_{0}}^{x}}
A_{x}(x^{\prime},y)dx^{\prime}+%
{\displaystyle\int\limits_{y_{0}}^{y}}
A_{y}(x_{0},y^{\prime})dy^{\prime}+\left\{
{\displaystyle\int\limits_{y_{0}}^{y}}
dy^{\prime}%
{\displaystyle\int\limits_{x_{0}}^{x}}
dx^{\prime}B(x^{\prime},y^{\prime})+g(x)\right\}  +f(y_{0})
\label{Lambda(x,y)1}%
\end{equation}
with\ $g(x)$ such that\ $\ \left\{
{\displaystyle\int\limits^{y}}
{\displaystyle\int\limits^{x}}
B+g(x)\right\}  \boldsymbol{:}$ $\mathsf{indep.}$ $\mathsf{of\ }x,$ and the
second is%

\begin{equation}
\Lambda(x,y)=\Lambda(x_{0},y_{0})+\int_{x_{0}}^{x}A_{x}(x^{\prime}%
,y_{0})dx^{\prime}+\int_{y_{0}}^{y}A_{y}(x,y^{\prime})dy^{\prime}+\left\{
{\Huge -}%
{\displaystyle\int\limits_{x_{0}}^{x}}
dx^{\prime}%
{\displaystyle\int\limits_{y_{0}}^{y}}
dy^{\prime}B(x^{\prime},y^{\prime})+h(y)\right\}  +\hat{h}(x_{0})
\label{Lambda(x,y)4}%
\end{equation}
with $h(y)$ such that\ $\ \left\{  {\Huge -}%
{\displaystyle\int\limits^{x}}
{\displaystyle\int\limits^{y}}
B+h(y)\right\}  \boldsymbol{:}$ $\mathsf{indep.}$ $\mathsf{of}\ y.$ These
results apply to cases where the particle goes through \textit{different}
perpendicular MFs (recall $B=B_{2}-B_{1}$) \textit{in spatial regions remote
to the observation point} $(x,y)$. One can again show that the 2 solutions are
equal for simple-connected space, and for multiple-connectivity the values of
multiplicities $f(y_{0})$ and $\hat{h}(x_{0})$ cancel out the nonlocalities
and reduce the above to the usual result of mere $A$-integrals along the 2
paths (i.e. two simple Dirac phases). For striped $B$-distributions, functions
$g(x)$ and $h(y)$ must be chosen in ways compatible with their above
conditions (as in the earlier $(x,t)$-cases)$\boldsymbol{;}$ by then taking
the \textit{difference} of (\ref{Lambda(x,y)1}) and (\ref{Lambda(x,y)4}) we
obtain that the \textquotedblleft AB phase\textquotedblright\ (originating
from the \textit{closed} line integral of $A$'s) is exactly cancelled by the
nonlocal term of MFs. This is reminiscent of the cancellation of phases
observed in the early experiments of Werner \& Brill\cite{WernerBrill} for
particles passing through a MF (a cancellation between the \textquotedblleft
AB phase\textquotedblright\ and the semiclassical phase picked up by the
trajectories), and our method seems to provide a natural
explanation$\boldsymbol{:}$ as our results are general (and for delocalized
states in simple-connected space they basically demonstrate the uniqueness of
$\Lambda$), they are also valid and applicable to states that describe
wavepackets in classical motion, as \textit{was} the case in Werner \& Brill's work.

The above cancellations can then be understood as a compatibility between the
AB fringe-displacement and the trajectory-deflection due to the Lorentz force
(i.e. the semiclassical phase picked up due to the optical path difference of
the two deflected trajectories \textit{exactly cancels} (is \textit{opposite
in sign} from) the AB phase picked up by the same trajectories due to the flux
that they enclose). This opposite sign seems to have been rather
unnoticed$\boldsymbol{:}$ In Feynman's Fig.15-8 \cite{Feynman}, or in
Felsager's Fig.2.16 \cite{Felsager}, classical trajectories are deflected
after passing through a strip of a MF placed on the right of a double-slit
apparatus. Both authors determine the semiclassical phase picked up by the
deflected trajectories and find it consistent with the AB phase. One can see
on closer inspection, however, that the two phases actually \textit{have
opposite signs }(see our own Fig.2 and the discussion that follows below,
where this is proved in detail). Similarly, in the very recent review of
Batelaan \& Tonomura\cite{Batelaan}, their Fig.2 shows wavefronts associated
to deflected classical trajectories where it is stated that \textquotedblleft
the phase shift calculated in terms of the Lorentz force is the same as that
predicted by the AB effect in terms of the vector potential\textquotedblright.
Once more, however, it turns out that the sign of the classical
phase-difference is really opposite to the sign of the AB phase (see proof
below). The phases are not equal as stated by the authors. And it turns out
that even \textquotedblleft electric analogs\textquotedblright\ of the above
cases also demonstrate this opposite-sign relationship (see proofs further
below). All the above examples can be viewed as a manifestation of the
cancellations that have been found in the present work for general quantum
states (but in those examples they are just special cases for
wavepacket-states in classical motion).

Let us give a brief elementary proof of the above claimed opposite
sign-relationships$\boldsymbol{:}$ Indeed, in our Fig.2, the \textquotedblleft
AB phase\textquotedblright\ due to the flux enclosed between the two classical
trajectories (of a particle of charge $q$) is
\begin{equation}
\Delta\varphi^{AB}=2\pi\frac{q}{e}\frac{\Phi}{\Phi_{0}}, \label{ABphase}%
\end{equation}
with $\Phi_{0}=\frac{hc}{e}$ the flux quantum, and $\Phi\thickapprox BWd$ the
enclosed flux between the two trajectories (for small trajectory-deflections),
with the deflection originating from the presence of the magnetic strip $B$
and the associated Lorentz forces. On the other hand, the semiclassical phase
difference between the same 2 classical trajectories is $\Delta\varphi
^{semi}=\frac{2\pi}{\lambda}\Delta l$, with $\lambda=\frac{h}{mv}$ being the
de Broglie wavelength (and $v$ being the speed of the particle, taken almost
constant (as usually done) due to the small deflections), and with $\Delta l$
being $\Delta l\thickapprox d\sin\theta\thickapprox d\frac{x_{c}}{L}$ ($x_{c}$
being the (displaced) position of the central fringe on the screen). We have
therefore
\begin{equation}
\Delta\varphi^{semi}=\frac{2\pi}{\lambda}d\frac{x_{c}}{L}. \label{Semiphase}%
\end{equation}
Now, the Lorentz force (exerted only during the passage through the thin
magnetic strip, hence only during a time interval $\Delta t=\frac{W}{v}$) has
a component parallel to the screen (let us call it $x$-component) that is
given by
\begin{equation}
F_{x}=\frac{q}{c}(\boldsymbol{v}\times\boldsymbol{B})_{x}=-\frac{q}%
{c}vB=-\frac{BWq}{c\frac{W}{v}}=-\frac{BWq}{c\Delta t} \label{Fx}%
\end{equation}
which shows that there is a change of kinematic momentum (parallel to the
screen) equal to $-\frac{BWq}{c},$ or, equivalently, a change of parallel speed%

\begin{equation}
\Delta v_{x}=-\frac{BWq}{mc} \label{Dvx}%
\end{equation}
which is the speed of the central fringe's motion (i.e. its displacement over
time along the screen). Although this has been caused by the presence of the
thin deflecting magnetic strip, this displacement is occuring uniformly during
a time interval $t=\frac{L}{v},$ and this time interval must satisfy%

\begin{equation}
\Delta v_{x}=\frac{x_{c}}{t} \label{Dvx2}%
\end{equation}
(as, for small displacements, the wps travel most of the time in uniform
motion, i.e. $\Delta t<<t$). We therefore have that the central fringe
displacement must be $x_{c}=\Delta v_{x}t=-\frac{BWq}{cm}\frac{L}{v},$ and
noting that $mv=\frac{h}{\lambda}$, we finally have%

\begin{equation}
x_{c}=-\frac{BWqL\lambda}{hc}. \label{xc}%
\end{equation}
By susbstituting (\ref{xc}) into (\ref{Semiphase}), the lengths $L$ and
$\lambda$ cancel out, and we finally have $\Delta\varphi^{semi}=-2\pi\frac
{q}{e}\frac{BWd}{\frac{hc}{e}},$ which with $\frac{hc}{e}=\Phi_{0}$ the flux
quantum, and $BWd\thickapprox\Phi$ the enclosed flux (always for small
trajectory-deflections) gives (through comparison with (\ref{ABphase})) our
final proof that%

\begin{equation}
\Delta\varphi^{semi}=-2\pi\frac{q}{e}\frac{\Phi}{\Phi_{0}}=-\Delta\varphi
^{AB}. \label{proof}%
\end{equation}
The \textquotedblleft electric analog\textquotedblright\ of the above exercise
is also outlined below, now with a homogeneous EF (pointing downwards
everywhere in space, but switched on for only a finite duration $T$) on the
right of a double-slit apparatus (see our Fig.3)$\boldsymbol{:}$ In this case
the electric Lorentz force $qE$ is exerted on the trajectories only during the
small time interval $\Delta t=T,$ which we take to be much shorter ($T<<t$)
than the time of travel $t=\frac{L}{v}$ (we now have a thin electric strip in
\textit{time} rather than the thin magnetic strip in space that we had
earlier). The electric type of AB phase is now%

\begin{equation}
\Delta\varphi^{AB}=-2\pi\frac{q}{e}\frac{cT\Delta V}{\Phi_{0}}, \label{ABele}%
\end{equation}
with $\Delta V$ being the electric potential difference between the two
trajectories, hence $\Delta V\thickapprox Ed$ (again for small
trajectory-deflections). On the other hand, the semiclassical phase difference
between the two trajectories is again given by (\ref{Semiphase}), but the
position $x_{c}$ of the central fringe must now be determined by the EF force
$qE\boldsymbol{:}$ The change of kinematic momentum (always parallel to the
screen) is now $qET$, hence the analog of (\ref{Dvx}) is now%

\begin{equation}
\Delta v_{x}=\frac{qET}{m} \label{Dvxnew}%
\end{equation}
which if combined with (\ref{Dvx2}) (that is obviously valid in this case as
well, again for small deflections, due to the $\Delta t=T<<t$), and always
with $t=\frac{L}{v}$, gives that the central fringe displacement must be
$x_{c}=\Delta v_{x}t=\frac{qET}{m}\frac{L}{v}$, and using again $mv=\frac
{h}{\lambda}$, we finally have the following analog of (\ref{xc})%

\begin{equation}
x_{c}=\frac{qETL\lambda}{h}. \label{xcnew}%
\end{equation}
By substituting (\ref{xcnew}) into (\ref{Semiphase}), the lengths $L$ and
$\lambda$ again cancel out, and we finally have $\Delta\varphi^{semi}=2\pi
d\frac{qETL\lambda}{h}=2\pi\frac{q}{e}\frac{EdcT}{\frac{hc}{e}}$, which with
$\frac{hc}{e}=\Phi_{0}$ the flux quantum, and through comparison with
(\ref{ABele}) leads once again to our final proof that%

\begin{equation}
\Delta\varphi^{semi}=-\Delta\varphi^{AB}. \label{proofnew}%
\end{equation}
We note therefore that even in the electric case, the semiclassical phase
difference (between two trajectories) picked up due to the Lorentz force
(exerted on them) is once again opposite to the electric AB phase picked up by
the same trajectories (due to the electric flux that they enclose).

We should point out once again, however, that although the above elementary
considerations apply to semiclassical motion of narrow wavepackets, in this
paper we have given \textit{a more general understanding of the above opposite
sign-relationships} that applies to general (even completely delocalized)
states, and that originates from our generalized Werner \& Brill cancellations.

In a slightly different vein, the cancellations that we found above give an
explanation of why certain classical arguments (invoking the past
$t$-dependent history of an experimental setup) seem to be successful in
giving at the end an explanation of AB effects (namely a phase consistent with
that of a static AB configuration). However, there is again an opposite sign
that seems to have been largely unnoticed in such arguments as well (i.e. in
Silverman\cite{Silverman}, where in his eq.(1.34) there should be an extra
minus sign).

Finally, on other shapes of $B$, see Fig.1b for an example of a
homogeneous$\ B$ distributed in a triangular shape (now the part of the
magnetic flux contained inside the \textquotedblleft observation
rectangle\textquotedblright\ depending on \textit{both} $x$ \textit{and} $y$).
It turns out that this flux can be written as a sum of separate $x$- and
$y$-contributions, and for an equilateral triangle of side $a$ we obtain that
proper functions (for the solutions (\ref{Lambda(x,y)1}) and
(\ref{Lambda(x,y)4})) are $\ g(x)=B\left(  \mathbf{-(}\sqrt{3}ax-\frac
{\sqrt{3}}{2}x^{2})+\frac{\sqrt{3}}{4}a^{2}\right)  $\ and $\ h(y)=B\left(
\mathbf{(}ay-\frac{y^{2}}{\sqrt{3}})-\frac{\sqrt{3}}{4}a^{2}\right)  $. These,
if substituted in (\ref{Lambda(x,y)1}) and (\ref{Lambda(x,y)4}), lead to new
and nontrivial nonlocal solutions (or, correspondingly, to \textit{nonlocal
phases} of wavefunctions). In cases of circularly shaped distributions (when
the enclosed flux is not separable) it is advantageous to solve the PDEs
directly in polar coordinates (for corresponding results see \cite{jphysa})
$-$ while for general shapes, one may need to first transform to an
appropriate coordinate system, and only then apply the above methodology (i.e.
strategy, for solving the resulting PDEs).

\section{Full (x,y,t)-case}

Finally, for the $t$-dependent 2-D case we have to solve$\ \frac
{\partial\Lambda}{\partial x}=A_{x},$\ $\frac{\partial\Lambda}{\partial
y}=A_{y},$ $-\frac{1}{c}\frac{\partial\Lambda}{\partial t}=\phi,$ in order to
see how the solutions \textit{combine} the spatial and temporal nonlocal
effects found above. We now have 3!=6 alternative routes to follow for
integrating the system and, at the end, 12 different results are derived,
where the $t$-propagation of $B$ and of $E_{x}$ and $E_{y}$ in all space is
nontrivially important. By leaving out all the long details\cite{jphysa} we
merely show one solution, where only $B(..,t_{0})$ appears (the $t$-dependence
of $B$ having already been incorporated in the behavior of $E_{x}$ and $E_{y}$
through Faraday's law), namely%

\[
\Lambda(x,y,t)=\Lambda(x_{0},y_{0},t_{0})+\int_{x_{0}}^{x}A_{x}(x^{\prime
},y_{0},t)dx^{\prime}+\int_{y_{0}}^{y}A_{y}(x,y^{\prime},t)dy^{\prime}-%
{\displaystyle\int\limits_{x_{0}}^{x}}
dx^{\prime}%
{\displaystyle\int\limits_{y_{0}}^{y}}
dy^{\prime}B(x^{\prime},y^{\prime},t_{0})+G(y,t_{0})-
\]

\[
-c%
{\displaystyle\int\limits_{t_{0}}^{t}}
\phi(x_{0},y_{0},t^{\prime})dt^{\prime}+c%
{\displaystyle\int\limits_{t_{0}}^{t}}
dt^{\prime}%
{\displaystyle\int\limits_{x_{0}}^{x}}
dx^{\prime}E_{x}(x^{\prime},y_{0},t^{\prime})+c%
{\displaystyle\int\limits_{t_{0}}^{t}}
dt^{\prime}%
{\displaystyle\int\limits_{y_{0}}^{y}}
dy^{\prime}E_{y}(x,y^{\prime},t^{\prime})+F(x,y)+f(x_{0},t_{0}),
\]
with conditions: $\left\{  G-%
{\displaystyle\int\limits^{x}}
{\displaystyle\int\limits^{y}}
B(x^{\prime},y^{\prime},t_{0})\right\}  \boldsymbol{:}\mathsf{indep.}$
$\mathsf{of}\ y,$ $\left\{  F+c%
{\displaystyle\int\limits^{t}}
{\displaystyle\int\limits^{x}}
E_{x}(x^{\prime},y,t^{\prime})\right\}  \boldsymbol{:}\mathsf{indep.}$
$\mathsf{of}\ x$,\ and $\left\{  F+c%
{\displaystyle\int\limits^{t}}
{\displaystyle\int\limits^{y}}
E_{y}(x,y^{\prime},t^{\prime})\right\}  \boldsymbol{:}\mathsf{indep.}$
$\mathsf{of}\ y.$ In the above, $f$ accounts for possible multiplicities at
$t_{0}$. This solution, together with its spatial \textquotedblleft
dual\textquotedblright\ $[$now with $\int_{x_{0}}^{x}A_{x}(x^{\prime
},y,t)dx^{\prime}+\int_{y_{0}}^{y}A_{y}(x_{0},y^{\prime},t)dy^{\prime}$
replacing the above $A$-terms, and with $c%
{\displaystyle\int\limits_{t_{0}}^{t}}
dt^{\prime}%
{\displaystyle\int\limits_{x_{0}}^{x}}
dx^{\prime}E_{x}(x^{\prime},y,t^{\prime})+c%
{\displaystyle\int\limits_{t_{0}}^{t}}
dt^{\prime}%
{\displaystyle\int\limits_{y_{0}}^{y}}
dy^{\prime}E_{y}(x_{0},y^{\prime},t^{\prime})$ replacing the above $E$-terms,
and with $G(y,t_{0})$ being replaced by a $\hat{G}(x,t_{0})$ that must
satisfy$\boldsymbol{:}$ $\left\{  \hat{G}+%
{\displaystyle\int\limits^{y}}
{\displaystyle\int\limits^{x}}
B(x^{\prime},y^{\prime},t_{0})\right\}  \boldsymbol{:}\mathsf{indep.}$
$\mathsf{of}\ x]$, are both crucial for the discussion of the
thought-experiment that follows$\boldsymbol{:}$ In \cite{vanKampen} van Kampen
considered a magnetic AB setup, but with an inaccessible magnetic flux that is
$t$-dependent$\boldsymbol{:}$ he envisaged turning on the flux very late, or
equivalently, observing the interference of the two wavepackets on a distant
screen very early, earlier than the time it takes light to travel the distance
to the screen (i.e. $t<\frac{L}{c}$), hence using the (instantaneous nature of
the) AB phase to transmit information (on the presence of a confined flux
somewhere in space) \textit{superluminally}. Indeed, the AB phase at any $t$
is determined by differences of $\frac{q}{\hbar c}\Lambda(\mathbf{r},t)$ with
$\Lambda(\mathbf{r},t)\sim\int_{\mathbf{r}_{0}}^{\mathbf{r}}\mathbf{A}%
(\mathbf{r}^{\prime},t)\boldsymbol{.}d\mathbf{r}^{\prime}$ (basically a
special case of (\ref{wrong})). However, if we use, instead, our results above
(that contain the additional nonlocal terms), it turns out\cite{jphysa} that,
for a spatially-confined flux $\Phi(t)$ and for $t<$ $\frac{L}{c}$, functions
$G$, $\hat{G}$ and $F$ can all be taken zero$\boldsymbol{\ }$(their conditions
are all satisfied), the point being that at instant $t$, the $\boldsymbol{E}%
$-field has not yet reached the spatial point $(x,y)$ of the screen $-$ a
generalization of the striped cases that we saw earlier but now to the case of
3 spatio-temporal variables (with now the spatial point $(x,y)$ being outside
the light-cone defined by $t$ (see Fig.4))$\boldsymbol{;}$ as the electric
flux is independent of the upper limits $x$ and $t$, this construction
rigorously gives $F=0$. Moreover, the AB multiplicities (at $t_{0}$) lead to
cancellation of the $B$-terms (always at $t_{0}$), with the final result
(after subtraction of the 2 solutions) being%

\begin{equation}
\Delta\Lambda(x,y,t)=%
{\displaystyle\oint}
\mathbf{A}(\mathbf{r}^{\prime},t)\boldsymbol{.}d\mathbf{r}^{\prime}+c%
{\displaystyle\int\limits_{t_{0}}^{t}}
dt^{\prime}%
{\displaystyle\oint}
\mathbf{E}(\mathbf{r}^{\prime},t^{\prime})\boldsymbol{.}d\mathbf{r}^{\prime}
\label{DeltaLambdaBrief}%
\end{equation}
which, with $%
{\displaystyle\oint}
\mathbf{A}(\mathbf{r}^{\prime},t)\boldsymbol{.}d\mathbf{r}^{\prime}=\Phi(t)$
\ the instantaneous enclosed magnetic flux and with the help of Faraday's law
$%
{\displaystyle\oint}
\mathbf{E}(\mathbf{r}^{\prime},t^{\prime})\boldsymbol{.}d\mathbf{r}^{\prime
}=-\frac{1}{c}\frac{d\Phi(t^{\prime})}{dt^{\prime}},$ \ gives%

\begin{equation}
\Delta\Lambda(x,y,t)=\Phi(t)-{\huge (}\Phi(t)-\Phi(t_{0}){\huge )}=\Phi
(t_{0}). \label{DeltaLambdaFinal}%
\end{equation}
Although $\Delta\Lambda$ is generally $t$-dependent, we obtain the intuitive
(causal) result that, for $t<\frac{L}{c}$ (i.e. if the physical information
has not yet reached the screen), the phase-difference turns out to be
$t$-independent, and leads to the magnetic Aharonov-Bohm\ phase that we
\textit{would} observe at $t_{0}$. \textit{The new nonlocal terms have
conspired in such a way as to exactly cancel the Causality-violating AB phase}
(that would be proportional to the instantaneous $\Phi(t)$). This gives a
resolution of the van Kampen \textquotedblleft paradox\textquotedblright%
\ within a canonical formulation, without using any vague electric AB argument
(as there is no multiple-connectivity in $(x,t)$-plane). An additional
physical element is that, for the above cancellation, it is not only the
$E$-fields but also the $t$-propagation of the $B$-fields (the full
\textquotedblleft radiation field\textquotedblright) that plays a
role\cite{jphysa}.

Use of the other 10 solutions can also address bound-state analogs (in
$t$-driven 1-D nanorings) or even \textquotedblleft electric\textquotedblright%
\ analogs of the van Kampen case$\boldsymbol{:}$ In Peshkin's
review\cite{Peshkin}, on the electric AB effect,$\boldsymbol{\ }$the author
correctly states \textquotedblleft One cannot wait for the electron to pass
and only later switch on the field to cause a physical
effect\textquotedblright. Although Peshkin uses his eq.(B.5) and (B.6) (based
on (\ref{wrong})), he carefully states that it is not the full solution;
actually, if we view it as an \textit{ansatz}, then it is understandable why
he needs to enforce a \textit{condition} (his eq.(B.8), and later (B.9)) on
the EF outside the cages (in order for certain (annoying) terms (resulting
from a minimal substitution due to the incorrect ansatz) to vanish and for
(B.5) to be a solution). But then he notes that the extra condition cannot
always be satisfied (hence (B.5) is not really the solution for all times),
drawing from this the above qualitatively correct conclusion on Causality. As
it turns out, our treatment gives exactly what Peshkin describes in words
(with the total \textquotedblleft radiation field\textquotedblright\ outside
the cages being once again crucial in recovering Causality), but in a direct
and fully quantitative manner, and with \textit{no ansatz} based on an
incorrect form. We should also point out that improper uses of simple Dirac
phases appear often in the literature$\boldsymbol{:}$ even in
Feynman\cite{Feynman} it is stated that the simple phase factor $\int%
^{x}\boldsymbol{A}\cdot d\mathbf{r}^{\prime}-c\int^{t}\phi dt^{\prime}$ is
valid even for dynamic fields; this is also explicitly stated in Erlichson's
review\cite{Erlichson} $-$ Silverman\cite{Silverman} being the only report
with a careful wording about (\ref{wrong}) being only restrictedly valid (for
$t$-indep. $\boldsymbol{A}$ and $\mathbf{r}$-indep. $\phi$), although even
there the nonlocal terms have been missed.

At the level of the basic Lagrangian $L(\mathbf{r},\mathbf{v},t)=\frac{1}%
{2}m\mathbf{v}^{2}+\frac{q}{c}\mathbf{v}\boldsymbol{.}\mathbf{A}%
(\mathbf{r},t)-q\phi(\mathbf{r},t)$ \ there are no fields present, and the
view holds in the literature\cite{BrownHome} that EFs or MFs cannot contribute
directly to the phase. This view originates from the path-integral treatments
widely used (where the Lagrangian determines directly the phases of
Propagators), but, nevertheless, our canonical treatment shows that fields
\textit{do} contribute nonlocally, and they are actually crucial in recovering
Relativistic Causality. Moreover, path-integral discussions\cite{Troudet} of
the van Kampen case use wave (retarded)-solutions for $\mathbf{A}$ (hence in
Lorenz gauge) and are incomplete; our results take advantage of the
retardation of \textit{fields} $\mathbf{E}$ and $\mathbf{B}$\textbf{ }(true in
\textit{any} gauge), and \textit{not }of potentials. In addition,
Troudet\cite{Troudet} correctly states that his path-integral treatment is
good for not highly-delocalized states in space, and that in case of
delocalization the proper treatment \textquotedblleft would be much more
complicated, and would require a much more complete analysis\textquotedblright%
. Such an analysis has actually been provided in the present work. It should
be added that the van Kampen \textquotedblleft paradox\textquotedblright%
\ seems to be still thought of as remarkable\cite{Compendium}. The present
work has provided a natural and general resolution, and most importantly,
through nonlocal (and Relativistically causal) propagation of wavefunction-phases.

On a broader significance of the new solutions we conclude that a causal
behavior may exist at the level of quantum mechanical phases, enforced by the
nonlocal terms (through the well-known causal behavior of fields). The
nonlocal terms found in this work at the level of $\Lambda$ reflect a causal
propagation of wavefunction-phases \textit{in the Schr\"{o}dinger picture} (at
least a part of them, the one containing the fields, that competes with the AB
types of phases containing the potentials). This nonlocality and Causality of
quantum phases is an entirely new concept (given the local nature but also the
nonrelativistic character of the SE) and deserves to be further explored.
Possible immediate applications would be in $t$-dependent slit-experiments
recently discussed using a completely different method (with modular variables
in the Heisenberg picture)\cite{Tollaksen}. It has been recently
noted\cite{He} that Physics cannot currently predict how we dynamically go
from the single-slit diffraction to the double-slit diffraction pattern
(whether it is in a gradual and causal manner or not). Application of our
nonlocal terms to such questions (i.e. by introducing scalar potentials on the
slits in a $t$-dependent way) provides a completely new formulation for
addressing causal issues of this type. Finally, one can always wonder what the
consequences of these new nonlocalities would be, if these were included in
other systems of High-Energy or Condensed Matter Physics \textit{with a gauge
structure}$\boldsymbol{;}$ alternatively, it is worth noting that, if $E$'s
were substituted by gravitational fields and $B$'s by Coriolis force fields
arising in non-inertial frames of reference, the above nonlocalities (and
their apparent causal nature) could possibly have an interesting story to tell
about quantum mechanical phase behavior in a Relativistic/Gravitational framework.

\section{Acknowledgements}

Students Kyriakos Kyriakou and Georgios Konstantinou of the Univ. of Cyprus
and Areg Ghazaryan of Yerevan State University are acknowledged for having
carefully reproduced all results. Georgios Konstantinou is also acknowledged
for having drawn Figures 2, 3 and 4. Dr. Cleopatra Christoforou of the
Department of Mathematics and Statistics of the Univ. of Cyprus is
acknowledged for a discussion concerning the mathematical method followed.

\textbf{Fig. 1.} (Color online)$\boldsymbol{:}$ Examples of
field-configurations (in simple-connected spacetime) where the nonlocal terms
are nonzero: (a) a strip in 1+1 spacetime, where the electric flux enclosed in
the \textquotedblleft observation rectangle\textquotedblright\ is dependent on
$t$ but independent of $x$; (b) a triangular distribution in 2-D space, where
the part of the magnetic flux inside the \textquotedblleft observation
rectangle\textquotedblright\ depends on both $x$ and $y$. The appropriate
choices for the corresponding functions $g(x)$ and $\hat{g}(t)$ for case (a),
or $g(x)$ and $h(y)$ for case (b), are given in the text.

\textbf{Figure 2.} (Color online)$\boldsymbol{:}$ The standard double-slit
apparatus with an additional strip of a perpendicular magnetic field $B$ of
width $W$ placed between the slit-region and the observation screen. The
deflection shown is for a negative charge $q$ (and in the text it is assumed
small, due to $W<<L$).

\textbf{Figure 3.} (Color online)$\boldsymbol{:}$ The analog of Fig.2 (again
for a negative charge $q$) but with an additional electric field parallel to
the observation screen that is turned on for a time interval $T$ (with $T<<t$,
and $t$ the time of travel).

\textbf{Figure 4.} (Color online)$\boldsymbol{:}$ The analog of paths of Fig.1
but now in 2+1 spacetime for the van Kampen thought-experiment, when the
instant of observation $t$ is so short that the physical information has not
yet reached the spatial point of observation $(x,y)$. The two solutions (that,
for wavepackets, have to be subtracted in order to give the phase difference
at $(x,y,t)$) are described in the text, and are here characterized through
their electric field $E$-line-integral behavior$\boldsymbol{:}$
\textquotedblleft electric field path (I)\textquotedblright\ (the red-arrow
route) denotes the \textquotedblleft dual\textquotedblright\ solution, and
\textquotedblleft electric field path (II)\textquotedblright\ (the green-arrow
route) denotes the \textquotedblleft primary\textquotedblright\ solution given
in the beginning of Section IV.

\end{document}